\documentclass[11pt, hidelinks, a4paper]{scrartcl}
  \usepackage{amsmath,amssymb}
  \usepackage{authblk}  
  \usepackage{caption}
  \usepackage{comment}
  \usepackage[top=30truemm,bottom=35truemm,left=30truemm,right=30truemm]{geometry}
  \usepackage{hyperref}
  \usepackage[section]{placeins} 
  \usepackage{tikz}  
	\usetikzlibrary{calc}
  \newcommand{\tn}[1]{\textnormal{#1}}
\title{\textbf{Inclusive Higgsstrahlung cross section measurements with the new reference sample method} }
\author{Jonas~Kunath\thanks{Presenter. Talk presented at the International Workshop on
    Future Linear Collider (LCWS2019), 28 October-1 November, 2019, Sendai, Japan. C19-10-28.} }
\author{Jean-Claude~Brient }
\affil{Laboratoire Leprince-Ringuet CNRS, \'Ecole Polytechnique,\hspace{2em} Institut Polytechnique de Paris, France}
\date{}
\begin{document}
\maketitle
\begin{abstract}
    An accurate determination of the Higgsstrahlung cross section is one of the main objectives
    at a future electron-positron collider.
    It allows for the only Higgs boson decay model independent measurement of the total Higgs width.
    Current results use the recoil mass shape method.
    That technique can be applied to Higgsstrahlung events with Z boson decays into muons,
    into electrons and, with reservations, into quarks.
    The samples built from Higgsstrahlung events with Z boson  decays into taus and neutrinos
    are not used in previous analyses.
    We present here a new method, the reference sample method.
    It extends the recoil mass method to be usable with the tau and neutrino samples as well.
    \par
    The extension promises a model independent determination of the inclusive Higgsstrahlung cross
    section with a $2.1-2.2 \%$ uncertainty from each of the two ILC polarization scenarios at $\sqrt{s}=250$~GeV
    with an integrated luminosity of $250~\tn{fb}^{-1}$.
    This represents an improvement of $20-30 \%$ on the accuracy from the application of the new approach
    without additional data collection.
    \end{abstract}
    \newpage
\section{Introduction}
  The Higgs boson production cross section at an electron-positron collider reaches its first significant
  peak at center-of-mass energies in the vicinity of 250~GeV.
  More than $95~\%$ of the Higgs bosons produced at this energy originate from Higgsstrahlung.
  \par
  The recoil mass technique enables an inclusive measurement of the Higgsstrahlung cross section~\cite{hRecoilShape}.
  By only utilizing the decay products of the recoiling Z boson the measurement is independent of the
  Higgs boson decay (model).
  The resulting estimation of the coupling between the Higgs boson and the Z boson can subsequently be exploited
  to break model dependence in estimations for all Higgs boson couplings.
  \par
  In this paper, we introduce a new approach that extends the recoil mass technique, making it applicable to events
  with $Z \rightarrow \tau^+ \tau^-$ and $Z \rightarrow \bar{\nu} \nu$ as well.
  We describe the new method and its implications on the uncertainty calculation.
  A proof of concept, based on generated events, is presented below.
\section{The reference sample method}
  The new method lifts the restriction of not being able to utilize any information that depends on the Higgs boson
  decay.
  The model independence is kept by extracting the efficiency of a Higgs boson decay dependent cut from observed
  events in a separate sample without any input from simulated events or theory predictions.
  \paragraph{}
  We build four samples: One for the hypothesis that the recoiling Z boson decayed invisibly into neutrinos and one
  for each of the hypothesis of the recoiling Z boson decaying into charged leptons of a specific flavor.
  \par
  For each of the charged lepton samples, we first identify groups of particle signatures in an event as candidates
  for a Z boson decaying into the respective charged lepton final state.
  A particle signature in such a group is tagged as a Z decay remnant.
  In accordance with the reference sample method, the quality of a Z boson candidate is assessed solely based on
  variables that can be defined from the Z decay remnants alone.
  Important quality features are an invariant mass close to the Z boson mass and a recoil mass close to the mass of
  the Higgs boson.
  Latter can be calculated from the knowledge on the design center-of-mass energy and the observed Z boson momentum:
  \begin{align}
    M_{\tn{rec}}^2 =\left( (\vec{0}, 250~\tn{GeV}) - (\vec{p}_Z, E_Z)\right)^2.
    \label{eq:mRec}
    \end{align}
  The events with an accepted quality are placed in the respective sample.
  As the decays in taus are harder to reconstruct and involve neutrinos, the quality requirement for this sample must
  be less tight.
  \par
  Following the same logic for the neutrino sample leads to identifying no particle signature as stemming from the Z
  decay.
  Clearly it is not possible to reject any events from this sample based on variables built from solely the Z decay
  remnants.
  Hence, every single event is a member of the neutrino sample.
  \paragraph{}
  For the events in all four samples it is possible to classify the particle signatures that pass the (kinematical)
  acceptance criteria.
  Based on whether it got tagged in the previous step a signature is classified as either stemming from the decay of
  a Z boson or not.
  We now discard the members of the Z boson class.
  \par
  What remains is a background event or a Higgsstrahlung event with the recoiling Z boson removed.
  By removing the recoiling Z boson we remove the difference between the Higgsstrahlung events in the four samples.
  The shape of the distribution of a variable built from the information remaining in the event is now independent
  of the sample for the signal events.\footnote{Due to dependence of the different Z boson selections on directional
  variables, this is not true for distributions depending on the direction of the Higgs boson.
  This can be overcome by a direction-depending reweighting of the distributions, or by simply avoiding
  direction-dependent variables.}
  \par
  While the background distributions depend on the sample, their shapes are well studied and can be taken from
  simulation with the appropriate uncertainties.
  \paragraph{}
  In deviation from the recoil mass method, we will now construct variables from the remaining part of the event.
  As in the signal case this remaining part are the particle signatures from the Higgs boson, this is henceforth called
  the Higgs-dependent selection.
  The educational example of such a variable that will be used in this paper is the number of charged hadron
  signatures in the event.
  Additional discriminators include the invariant mass of the remaining event objects or the number of electrons.
  \par
  As stated above, it is not allowed to take the efficiency of such a selection from anywhere but the experiment itself.
  We will thus employ one or multiple of the samples to extract the Higgs-dependent selection efficiency.
  Since a high sample purity is needed for this task, it should be performed on the muon or electron sample.
  A sample that is given this role will be named \emph{reference sample}.
  \par
  For each of the remaining samples a Higgs-dependent selection is performed to improve on the purity.
  Note that this is additional selection step is in particular inevitable for the neutrino sample.
  Before the step every event recorded in the detector is contained in the neutrino sample.
  After this additional selection, the remaining part of the sample is called the \emph{counting sample}.
  It will be used for an estimation of the Higgsstrahlung cross section.
  The selection efficiency for a counting sample is extracted from the reference sample by applying the same
  selection to it.
\section{Simulation}
  In this study, we use simulated events before the detector reconstruction.
  They were prepared for the Detailed Baseline Design (DBD) of the International Large Detector (ILD)
  concept~\cite{ILD_DBD} with ILCSoft v01-16-p10$\_$250~\cite{ILCSoft}.
  The event samples were generated using WHIZARD version 1.95~\cite{whizard,omega}.
  The fragmentation and hadronization of final-state quarks and gluons was performed with PYTHIA 6.4~\cite{pythia}.
  The tau lepton decays were simulated using TAUOLA~\cite{tauola}.
  \par
  The events are scaled to an integrated luminosity of $250~\tn{fb}^{-1}$ at center-of-mass energy
  $\sqrt{s}=250~\tn{GeV}$ for each of the two polarization scenarios.
  These scenarios are a right polarized run
  ($80~\%$ right-handed electron beam and $30~\%$ left-handed positron beam) and a left polarized run
  ($80~\%$ left-handed electron beam and $30~\%$ right-handed positron beam).
  \par
  Neutrinos, particles with $\left| \tn{cos} (\theta)\right| > 0.995$, charged particles with
  $p_{\tn{T}} < 0.105~\tn{GeV}$, photons with $E_{\gamma} < 0.2~\tn{GeV}$
  and neutral hadrons with $E < 0.5~\tn{GeV}$ are discarded for this study.
\section{Uncertainty calculation}
  For an estimation of the Higgsstrahlung cross section from a counting sample we need an estimation on the number
  of signal events in the counting sample ($N_C$, from the event count in this sample, $D_C$,
  and the expected background).
  Additionally the signal efficiency of the Higgs-dependent selection as well as the signal efficiency of the
  event selection from the remnants of the recoiling Z boson have to be known.
  The uncertainties on the branching ratio of the Z boson into the respective final state and the integrated
  luminosity are negligible.
  Remember that the Z-dependent selection efficiency is calculated from a generated Monte Carlo sample.
  Its uncertainty was shown to be small compared to the expected statistical uncertainty on the cross section
  determination for any ILC running scenario~\cite{hRecoilIndependence}.
  \begin{align}
    \sigma_{HZ} &= \frac{N_{HZ}}{L} = \frac{N_C}{BR(Z\rightarrow l \bar{l})\epsilon_Z \epsilon_H L} \\
    \frac{ \Delta\sigma_{HZ} }{ \sigma_{HZ} }
    &\approx \sqrt{ \left( \frac{ \Delta N_C }{ N_C} \right)^2   +    \left( \frac{ \Delta \epsilon_H }{ \epsilon_H} \right)^2 } \\&
    = \sqrt{  \frac{ D_C }{ (N_C)^2 } + \frac{ D_R }{ (N_R)^2 } - \frac{ 2 D_R^C}{ N_R^C N_R } + \frac{ D_R^C}{ (N_R^C)^2 } }.
    \label{eq:effErr}
    \end{align}
  The first term in equation \ref{eq:effErr} is the Poissonian event count uncertainty for the number of
  signal events in the counting sample.
  The remaining three terms emerge from the binomial uncertainty on the Higgs-dependent selection efficiency.
  They are uncorrelated to the first term as they are based on a different sample, the reference sample.
  Two quantities must be obtained from the reference sample for each counting sample: The total number of events in the
  reference sample, $N_R$ and the number of events that additionally are selected by the Higgs-dependent selection of
  this counting sample, $N_R^C$.
\section{Analysis}
  \begin{figure}[ht]
    \centering
    \begin{tikzpicture}
      \node (img1) {\includegraphics[width=0.95\textwidth, keepaspectratio]{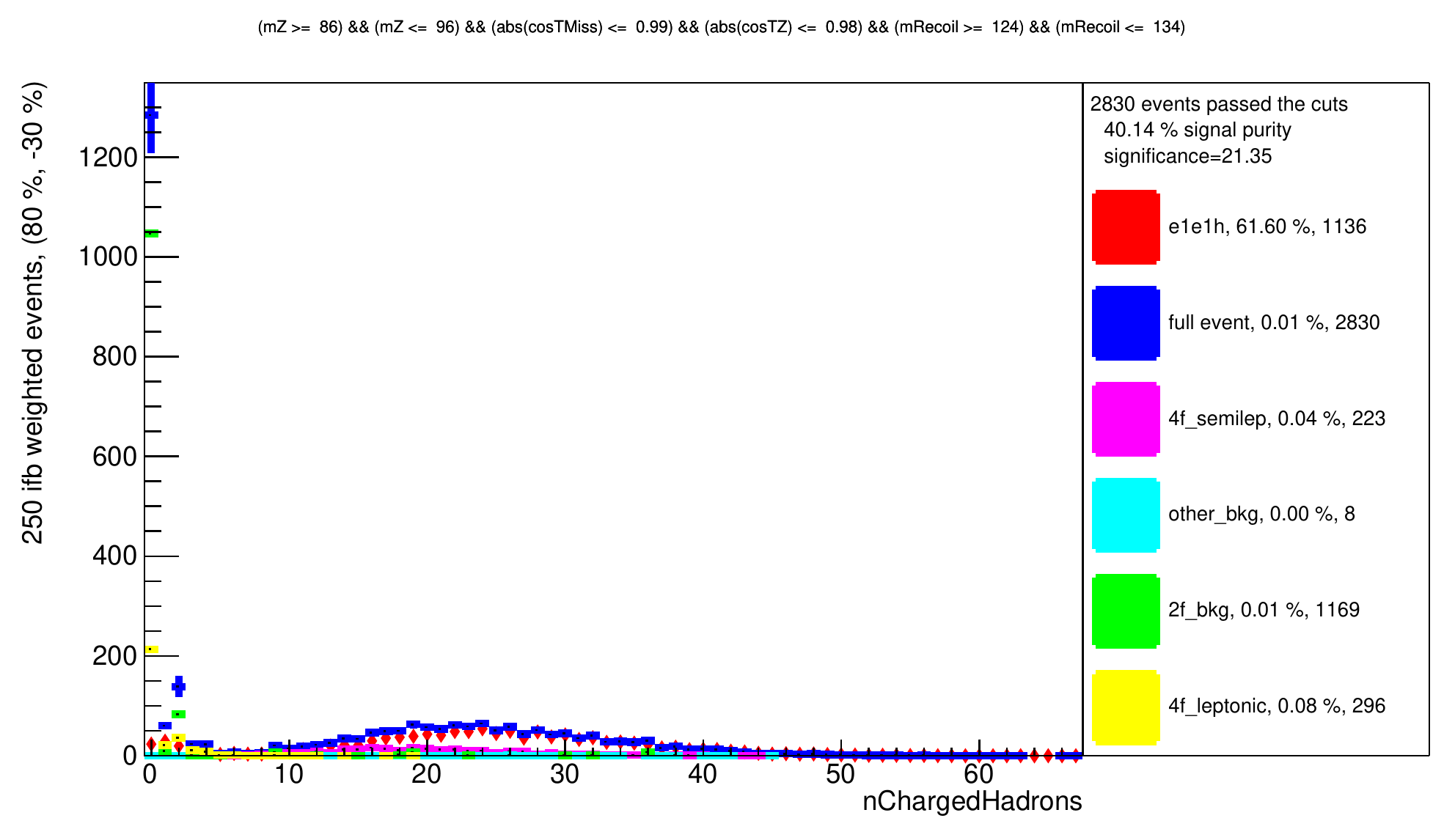}};
      \node(smallCenter) at (
          $0.3*(img1.north west) + 0.3*(img1.north east)+0.2*(img1.south west) + 0.2*(img1.south east)$) {};
      \node (img2.center) at (smallCenter) {
          \includegraphics[width=0.45\textwidth, keepaspectratio]{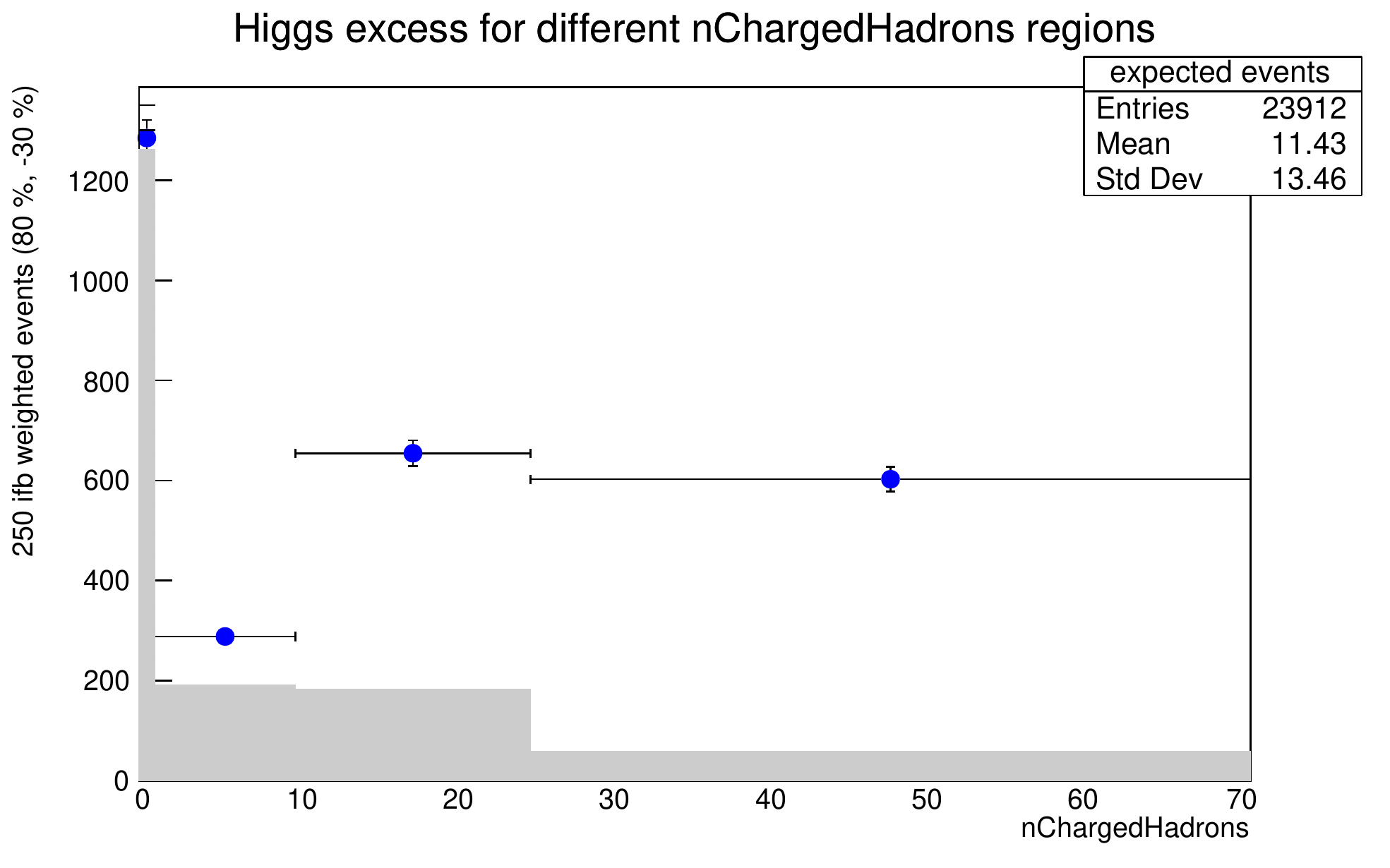}};
      \end{tikzpicture}
    \caption{Distribution in the number of charged hadrons for the $(+80\%, -30\%)$ polarized electron sample with
      selection cuts as specified in the title.
      The larger graph gives the overall event count per bin (blue) with its statistical uncertainty as well
      as the different process group contributions to each bin.
      The small plot has the same events in only four bins.
      Its histogram is built from the number of background events per bin.
      The markers indicate the expected number of observed events with uncertainty.
      \emph{Entries} is the number of Monte Carlo events that pass the selection and are used in the
      plot before event weighting.}
    \label{fig:nHadrDistrElectron}
    \end{figure}
  \begin{figure}[ht]
    \centering
    \includegraphics[width=0.95\textwidth, keepaspectratio]{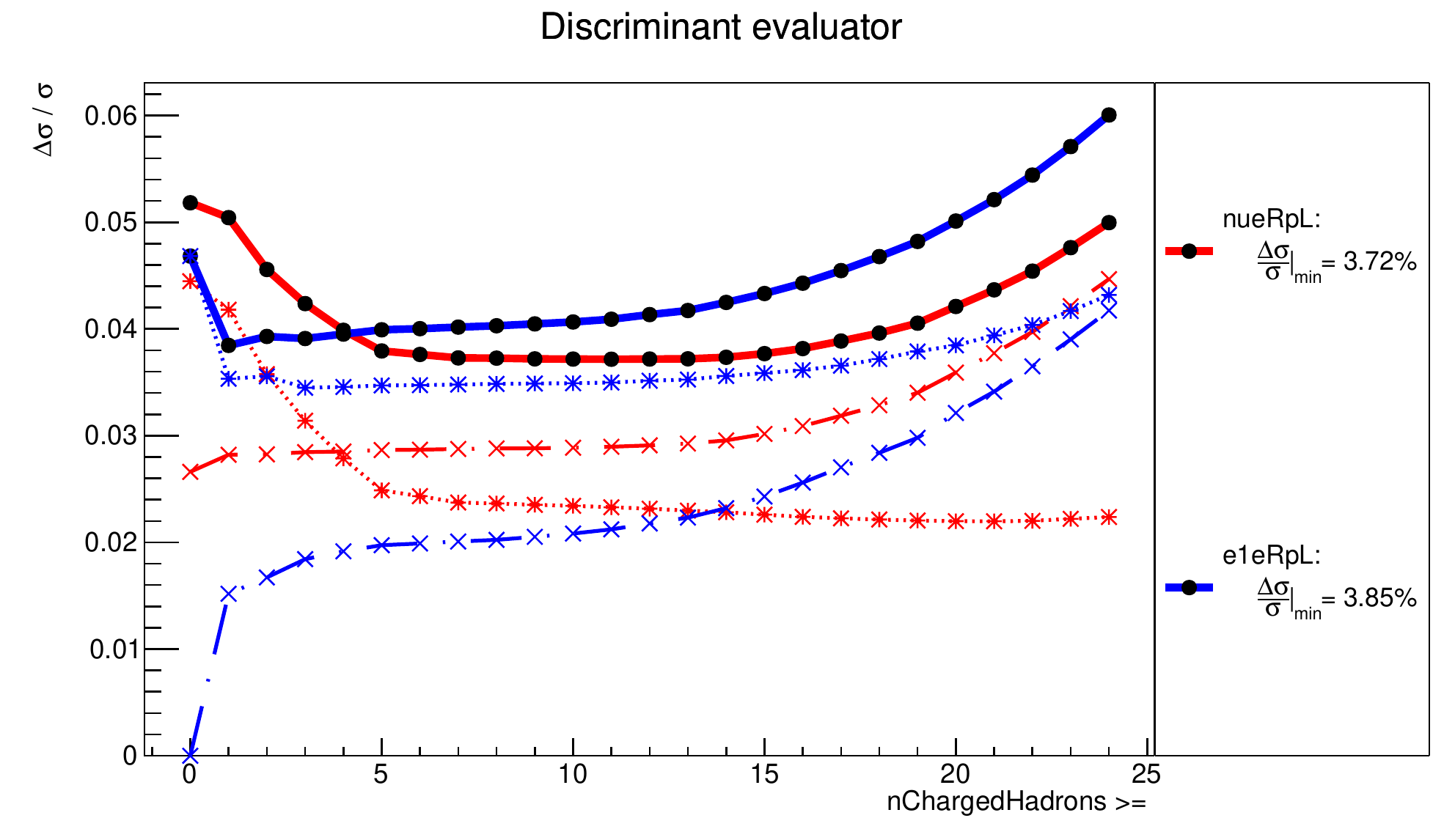}
    \caption{Effect of a cut on the number of charged hadrons on the relative cross section uncertainty.
      The displayed samples are the neutrino and electron sample of the right polarized run scenario.
      The contributions on the uncertainty from the estimation of the number of events in the counting sample
      (dotted line) and the Higgs-dependent selection efficiency are given as well.
      The best achievable uncertainty for each sample is cited in the legend.}
    \label{fig:partialErr}
    \end{figure}
  The electron sample before a Higgs-dependent selection is shown in figure \ref{fig:nHadrDistrElectron}.
  It could be used as a reference sample by applying the Higgs-dependent selections defined for some other
  counting samples onto it and counting the respective numbers of remaining elements.
  \par
  Alternatively, we can apply a Higgs dependent selection onto this sample and have the remaining elements
  form a counting sample.
  The signal events tend to have a higher number of charged hadron signatures.
  Especially requiring at least one charged hadron in the event removes  a high number of background events
  without losing many signal events.
  \par
  The trade-off between the two dominant sources of uncertainty when choosing the strength of the
  Higgs-dependent selection is emphasized in figure~\ref{fig:partialErr}.
  As the neutrino sample has additional Higgs-dependent selection criteria applied to it, the
  efficiency uncertainty is already non-zero without a cut on the number of charged hadrons.
  \par
  Additional information is given in the appendix.
  The uncertainty on the Higgsstrahlung cross section as a function of the number of charged hadrons for all
  six considered counting samples is described in figure~\ref{fig:allPols}.
  The exact selection criteria chosen are summarized in tables~\ref{tab:HiggsCuts} and~\ref{tab:ZCuts}.
  %
  \paragraph{}
  The relative uncertainties on the Higgsstrahlung cross section for the presented analyses with the
  new reference sample method are summarized in table~\ref{tab:errorSummary}.
  Results obtained with the recoil mass shape method~\cite{hRecoilShape} at the same collider conditions
  are cited for comparison.
  Note however that those results were obtained after a full ILD detector reconstruction.
  \par
  Only electron and muon sample can be used with the pure recoil mass method.
  As the two recoil mass shape analyses use independent data, and the dominating uncertainty is statistical,
  their uncertainties are taken to be uncorrelated.
  The uncertainties of the reference sample estimators are correlated.
  They are all based on a muon reference sample.
  The combined uncertainties are approximated with toy studies.
  Since the reference sample method is envisaged as an extension of the recoil mass method and not
  a replacement, it is natural to try to combine the results.
  \par
  Both methods build an estimator from an electron sample.
  It seems to be favorable to use the electron sample from the established recoil mass shape method,
  as its uncertainty is uncorrelated to the uncertainties from tau and neutrino sample.
  Since this means that the electron sample is not used as a counting sample in the reference sample method,
  it could now be added to the reference sample.
  As the signal count in the reference sample would then be doubled, we expect the uncertainty component
  from the efficiency to decrease significantly.
  \par
  Independent of further improvement opportunities the reference sample method reveals the potential for
  a significant improvement on the Higgsstrahlung cross section.
  \begin{table}[ht]
    \centering
    \begin{tabular}{l|cc|cc}
        250~GeV & $e^-_L e^+_R$ rec. shape & $e^-_L e^+_R$ new &
            $e^-_R e^+_L$ rec. shape & $e^-_R e^+_L$ new  \\ \hline
        $H \mu^+ \mu^-$ & $3.2\%$ & x & $3.6\%$ & x \\
        $H e^+ e^-$ & $4.0\%$ & $3.9\%$ & $4.7\%$ & $3.8\%$ \\
        $H \tau^+ \tau^-$ & x & $4.6\%$ & x & $4.9\%$ \\
        $H \nu \bar{\nu}$ & x & $4.2\%$ & x & $3.7\%$ \\ \hline
        combined & $2.5\%$ &  $3.0\%$ & $2.9\%$ &  $2.8\%$ \\
        $\mu$ from rec. & \multicolumn{2}{c|}{$2.4\%$} & \multicolumn{2}{c}{$2.2\%$} \\
        $\mu$ \&  $e$ from rec. & \multicolumn{2}{c|}{$2.1\%$} & \multicolumn{2}{c}{$2.2\%$} \\
        \end{tabular}
    \caption{Summary table of the relative uncertainties on the Higgsstrahlung cross section with the
      established recoil mass shape approach~\cite{hRecoilShape} and with the new reference sample method,
      introduced here.
      The results are compared and combined for the right and left polarized $250~\tn{fb}^{-1}$ run scenarios
      of the ILC at 250 GeV center-of-mass energy.
      A sample that is not used by a method is denoted with \emph{x}.
      The \emph{combined} row shows the uncertainty for combining the measurements from the same method.
      If applicable, the correlation of uncertainties is taken into account.
      The last two rows show the prospect of combining the results of both methods.
      The electron sample result can be taken from either method.}
    \label{tab:errorSummary}
    \end{table}
\section{Summary}
  Extending the recoil mass method with the presented reference sample method promises significant
  improvements on the inclusive Higgsstrahlung cross section measurement at an electron-positron
  collider at 250~GeV center-of-mass energy.
  By extracting the efficiency of a Higgs-dependent selection data-driven from a part of the Higgsstrahlung sample
  we can apply a Higgs-dependent selection to the rest of our sample without losing model independence.
  \par
  The study with simulated data suggests that $20-30~\%$ improvement on the uncertainty are possible.
  We will carry on with a full reconstruction study and optimized selection criteria.
  As indicated by this study, the Higgsstrahlung events with the Z boson decaying to an electron pair
  will be added to the reference sample.
\section*{Acknowledgements} 
  The authors are thankful to the ILD collaboration for the permitted use of the generator samples in this study.
\providecommand{\href}[2]{#2}\begingroup\raggedright
\endgroup 
\newpage
\appendix
\section*{Appendix}
  \begin{figure}[ht]
    \centering
    \includegraphics[width=0.95\textwidth, keepaspectratio]{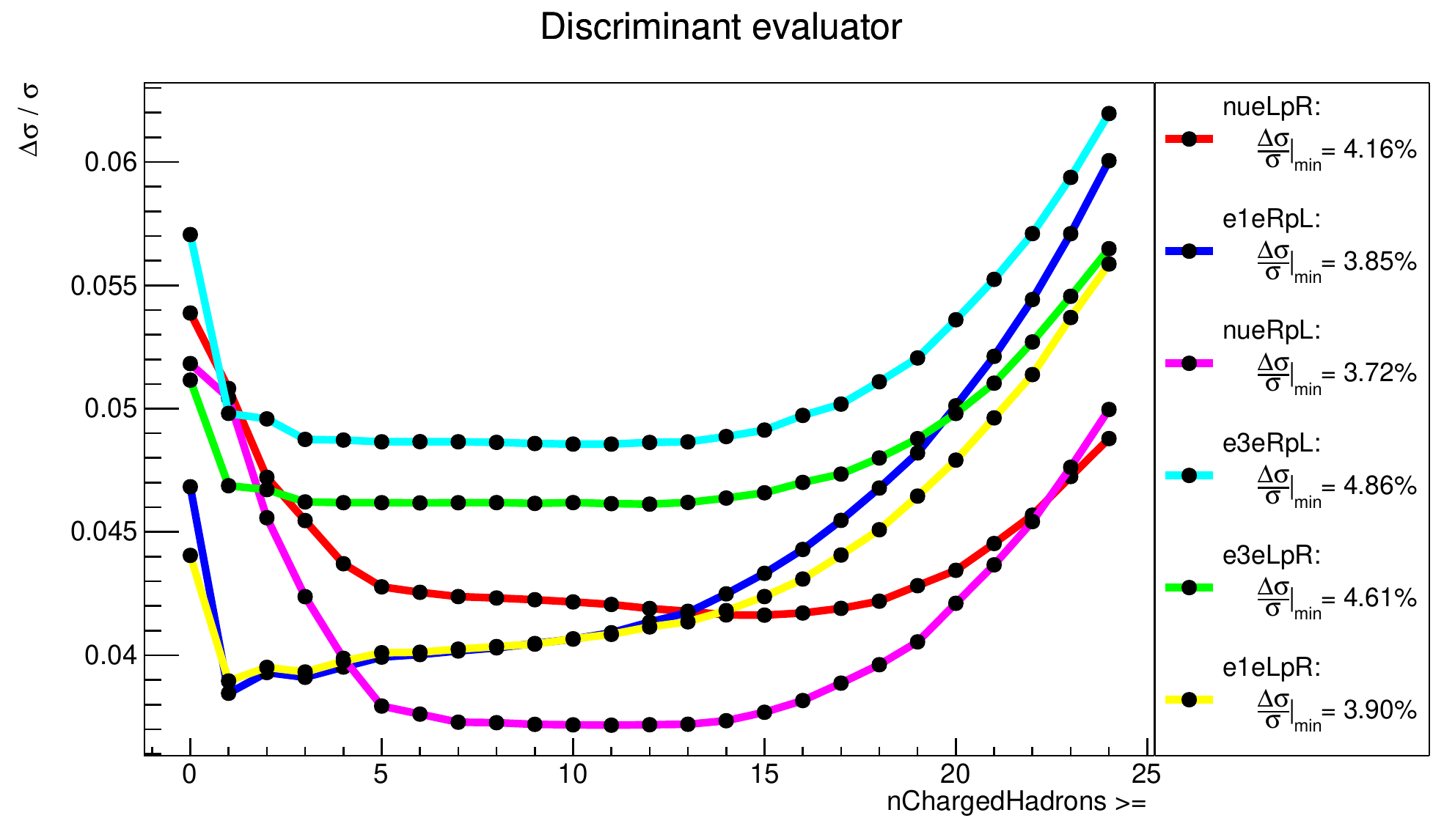}
    \caption{Effect of a cut on the number of charged hadrons on the relative cross section uncertainty.
      Neutrino, electron and tau sample are displayed for both the right (eRpL) and the left (eLpR) polarized
      run scenario.
      The best achievable uncertainty for each sample is cited in the legend.}
      \label{fig:allPols}
    \end{figure}
  \begin{table}\centering
    \begin{tabular}{l|ccc}
        Higgs-dependent cuts & $M_H^{\textnormal{recoil}} $ & $M_H $ & $\#$ ch. hadrons\\ \hline
        $Z \rightarrow e^+e^-$, eLpR & x & x & $\geq 1$\\
        $Z \rightarrow \tau^+\tau^-$, eLpR & $\geq 83$ & $\in \left[106, 130\right]$ & $\geq 10$ \\
        $Z \rightarrow \bar{\nu} \nu$, \hspace{1em} eLpR & $\in \left[87, 130\right]$ &
            $\in \left[104, 128\right]$ & $\geq 15$ \\ \hline
        $Z \rightarrow e^+e^-$, eRpL & x & x & $\geq 1$\\
        $Z \rightarrow \tau^+\tau^-$, eRpL & $\geq 80$ & $\in \left[103, 132\right]$ & $\geq 10$ \\
        $Z \rightarrow \bar{\nu} \nu$, \hspace{1em} eRpL & $\in \left[86, 135\right]$ &
            $\in \left[97, 130\right]$ & $\geq 10$ \\
        \end{tabular}
    \caption{The Higgs-dependent cuts applied to the counting samples in the presented analysis.
      The lines below the dividing line show those cuts utilized for the right polarized
      $(+80\%, -30\%)$ run scenario.
      Variables with value x are not utilized in that sample.}
    \label{tab:HiggsCuts}
    \end{table}
  \begin{table}\centering
    \begin{tabular}{l|ccccc}
        recoiling Z cuts & $M_Z$ & $M_{\textnormal{recoil}}$ &
            $\left| \textnormal{cos}(\theta_{\textnormal{miss}})\right|$ &
            $\left| \textnormal{cos}(\theta_Z)\right|$ \\ \hline
        $Z \rightarrow e^+e^-$, eLpR & $\in \left[88, 94\right]$
            & $\in \left[124, 127\right]$ &  $\leq 0.93$ & $\leq 0.99$\\
        \hspace{2em}$\mu^+\mu^-$, for $Z\rightarrow e^+e^-$& $\in \left[86, 96\right]$
            & $\in \left[124, 130\right]$ & $\leq 0.98$ & $\leq 0.99$\\
        $Z \rightarrow \tau^+\tau^-$, eLpR & x
            & x &  x & $\leq 0.99$\\
        \hspace{2em}$\mu^+\mu^-$, for $Z\rightarrow \tau^+\tau^-$& $\in \left[88, 94\right]$
            & $\in \left[124, 127\right]$ & $\leq 0.93$ & $\leq 0.99$\\
        $Z \rightarrow \bar{\nu} \nu$, eLpR & x
            & x &  x & $\leq 0.99$\\
        \hspace{2em}$\mu^+\mu^-$, for $Z\rightarrow \bar{\nu} \nu$& $\in \left[88, 94\right]$
            & $\in \left[124, 127\right]$ & $\leq 0.93$ & $\leq 0.99$\\ \hline
        $Z \rightarrow e^+e^-$, eRpL & $\in \left[88, 94\right]$
            & $\in \left[124, 127\right]$ &  $\leq 0.93$ & $\leq 0.99$\\
        \hspace{2em}$\mu^+\mu^-$, for $Z\rightarrow e^+e^-$& $\in \left[86, 96\right]$
            & $\in \left[124, 134\right]$ & $\leq 0.98$ & $\leq 0.99$\\
        $Z \rightarrow \tau^+\tau^-$, eRpL & x
            & x &  x & $\leq 0.99$\\
        \hspace{2em}$\mu^+\mu^-$, for $Z\rightarrow \tau^+\tau^-$& $\in \left[88, 94\right]$
            & $\in \left[124, 127\right]$ & $\leq 0.93$ & $\leq 0.99$\\
        $Z \rightarrow \bar{\nu} \nu$, eRpL & x
            & x &  x & $\leq 0.99$\\
        \hspace{2em}$\mu^+\mu^-$, for $Z\rightarrow \bar{\nu} \nu$& $\in \left[88, 94\right]$
            & $\in \left[124, 127\right]$ & $\leq 0.93$ & $\leq 0.99$\\
        \end{tabular}
    \caption{The cuts on the recoiling Z boson in the presented analysis.
      The analysis uses the muon sample as its reference sample.
      The selection producing the muon sample is optimized to each of the six analyses.
      The lines below the dividing line show those cuts utilized for the right polarized $(+80\%, -30\%)$ run scenario.
      Variables with value x are not utilized in that sample.
      The tau sample was treated differently from how it will be treated in the following full reconstruction study.
      Instead of reconstruction a tau and obtaining restricting quality cuts, we selected as tau candidate
      events any events with a Monte Carlo truth tau pair and subsequently removed all signatures originating
      from the tau pair from the event.}
    \label{tab:ZCuts}
    \end{table}
\end{document}